\documentclass[aip,cha,reprint,numerical, nofootinbib, groupedaddress]{revtex4-1}

\usepackage{color}
\usepackage{amsmath, amsthm, amssymb}
\usepackage{graphicx}
\usepackage{listings}
\usepackage{subfigure}

\begin{document}
	
\title{Nonlinear time-series analysis revisited} \thanks{EB
  thanks the Max-Planck-Institut fur Physik komplexer Systeme for
  hosting the visit during which this paper was written.}

	\author{Elizabeth Bradley}
		\email{lizb@cs.colorado.edu}	
		\affiliation{Department of Computer Science, University of Colorado, Boulder CO 80309-0430 USA}
		\affiliation{Santa Fe Institute, Santa Fe, New Mexico 87501, USA}

	\author{Holger Kantz}
		\email{kantz@pks.mpg.de}	
		\affiliation{Max Planck Institute for the Physics of Complex Systems, Noethnitzer Str. 38
D 01187 Dresden
Germany}

\begin{abstract}

In 1980 and 1981, two pioneering papers laid the foundation for what
became known as {\sl nonlinear time-series analysis}: the analysis of
observed data---typically univariate---via dynamical systems theory.
Based on the concept of state-space reconstruction, this set of
methods allows us to compute characteristic quantities such as
Lyapunov exponents and fractal dimensions, to predict the future
course of the time series, and even to reconstruct the equations of
motion in some cases.  In practice, however, there are a number of
issues that restrict the power of this approach: whether the signal
accurately and thoroughly samples the dynamics, for instance, and
whether it contains noise.  Moreover, the numerical algorithms that we
use to instantiate these ideas are not perfect; they involve
approximations, scale parameters, and finite-precision arithmetic,
among other things.  Even so, nonlinear time-series analysis has been
used to great advantage on thousands of real and synthetic data sets
from a wide variety of systems ranging from roulette wheels to lasers
to the human heart.  Even in cases where the data do not meet the
mathematical or algorithmic requirements to assure full topological
conjugacy, the results of nonlinear time-series analysis can be
helpful in understanding, characterizing, and predicting dynamical
systems.

\end{abstract}

\maketitle

\begin{quotation} 

Nonlinear time-series analysis comprises a set of methods that extract
dynamical information about the succession of values in a data set.
This framework relies critically on the concept of reconstruction of
the state space of the system from which the data are sampled.  The
foundations for this approach were laid around 1980, when
deterministic chaos became a popular field of research and scientists
were looking for evidence of chaos in natural and laboratory systems.
One of the first---and still most spectacular---applications was the
prediction of the path of a ball on a roulette wheel, which nicely
demonstrated the power of these methods.  Since then, nonlinear
time-series analysis has left this narrow niche and moved into much
broader use across all branches of science and engineering, as well as
social science, the humanities, and beyond.


\end{quotation}

\section{Why nonlinear time series analysis?} \label{sec:intro}

The goal of time-series analysis is to learn about the dynamics behind
some observed time-ordered data.  Early approaches to this employed
linear stochastic models---more precisely, autoregressive (AR) and
moving average (MA) models\cite{BoxJenkins}.  These stationary
Gaussian stochastic processes are fully characterized by their
two-point auto-correlation function
\begin{equation}
c(\tau)= \frac{\langle(x_t - x_{t+\tau})^2\rangle}{\langle x_t^2\rangle}
\end{equation}
or by their power spectrum, respectively.  There are many data sets
where this type of analysis leads to a good characterization, such
as temperature anomalies: differences between the daily (maximum,
mean, minimum) temperature at a given place and the many-year average
of that quantity for the corresponding calendar day.  Data of this
type possess an almost Gaussian distribution with an almost
exponentially decaying auto-correlation function; typically the null
hypothesis that they are generated by an AR(1) process cannot be
rejected easily on the basis of observed data.

Of course, we know that temperatures can be predicted much more
accurately by high-dimensional physics-based models of the atmosphere
than by AR(1) models.  That scalar temperature data look like AR data
comes from the projection of dynamics in a high-dimensional state
space onto a single quantity.  This illustrates that, depending on
one's point of view and one's access to a system's variables, the very
same system might appear to have very different complexity.

As in any other analysis, the choice of a specific time-series
analysis method requires justification by some hypothesis about the
appropriate {\sl data model}.  Time-series analysis is essentially
data compression: we compute a few characteristic numbers from a large
sample of data.  This reduced information can only enhance our
knowledge about the underlying system if we can interpret it, and it
becomes interpretable through the fact that the chosen number has some
specific meaning within some model framework.  If the data do not stem
from the appropriate model class, the chosen quantity might not make
much sense, even if we can compute its numerical value on the given
data set using some numerical algorithm.  An illustrative example is
the computation of the mean and the variance of some sample: we know
how to do this, but are these two numbers always meaningful?  If the
hypothesis is well justified that the observed data are a sample from
a Gaussian distribution, then these numbers characterize it completely
and there is nothing else to compute.  If, on the other hand, the data
stem from a bimodal distribution, then the (still well defined) mean
value is very atypical and the variance is not the most interesting
feature.



The collection of ideas and techniques known as nonlinear time-series
analysis can be extremely effective when the data model is
deterministic dynamics in some state space.  This analysis framework
allows us to solve an inverse problem of considerable complexity: from
data, we can infer properties of the invariant measure of some hidden
dynamical system.  In the best case, we can even determine equations
of motion.  And, if the underlying system is deterministic and low
dimensional, this analysis framework brings out the relationships
between geometry (fractal dimension), instability (Lyapunov
exponents), and unpredictability (K-S entropy), which is a beautiful
theoretical result from ergodic theory.  Of course, the assumption of
determinism makes these methods largely unsuitable for characterizing
stochastic aspects of data.  Anomalous diffusion, as first observed in
Hurst's study of time-series data of the river Nile\cite{Hurst}, is
nowadays studied using detrended fluctuation analysis \cite{Pengetal};
behavior like this is a signature of both nonlinearity and
non-Gaussianity in the underlying stochastic process.

In this article, we want to describe---without too much detail or any
attempt at a comprehensive bibliography---the ideas and concepts of
nonlinear time-series analysis, to give a fair account of their
usefulness, and to offer some perspectives for the future.  Readers
who want to enter this subject more deeply should consult one of the
many comprehensive review articles or useful monographs on this topic,
such as \cite{Abarbanel,KantzSchreiber}.

%

\section{The Basics} \label{sec:basics}

State-space reconstruction is the foundation of nonlinear time-series
analysis.  This quite remarkable result, which was first proposed by
Packard {\sl et al.} in 1979 and 1980 \cite{jpcthesis,packard80} and
formalized by Takens soon thereafter \cite{takens}, allows one to
reconstruct the full dynamics of a complicated nonlinear system from a
single time series, in principle.  
The reconstruction is not, of course, identical
to the internal dynamics, or this procedure would amount to a general
solution to control theory's {\sl observer problem}: how to identify
all of the internal state variables of a system and infer their values
from the signals that {\sl can} be observed.  Even so, these
reconstructions---if done right---can still be extremely useful
because they are guaranteed to be topologically identical to the full
dynamics.  And since many important properties of dynamical systems
are invariant under diffeomorphism, this means that conclusions drawn
about the reconstructed dynamics also hold for the true dynamics of
the system.

\subsection{Delay-coordinate embedding}
\label{sec:dce}

\begin{figure*}
    \centering
    \includegraphics[width=1.3\columnwidth]{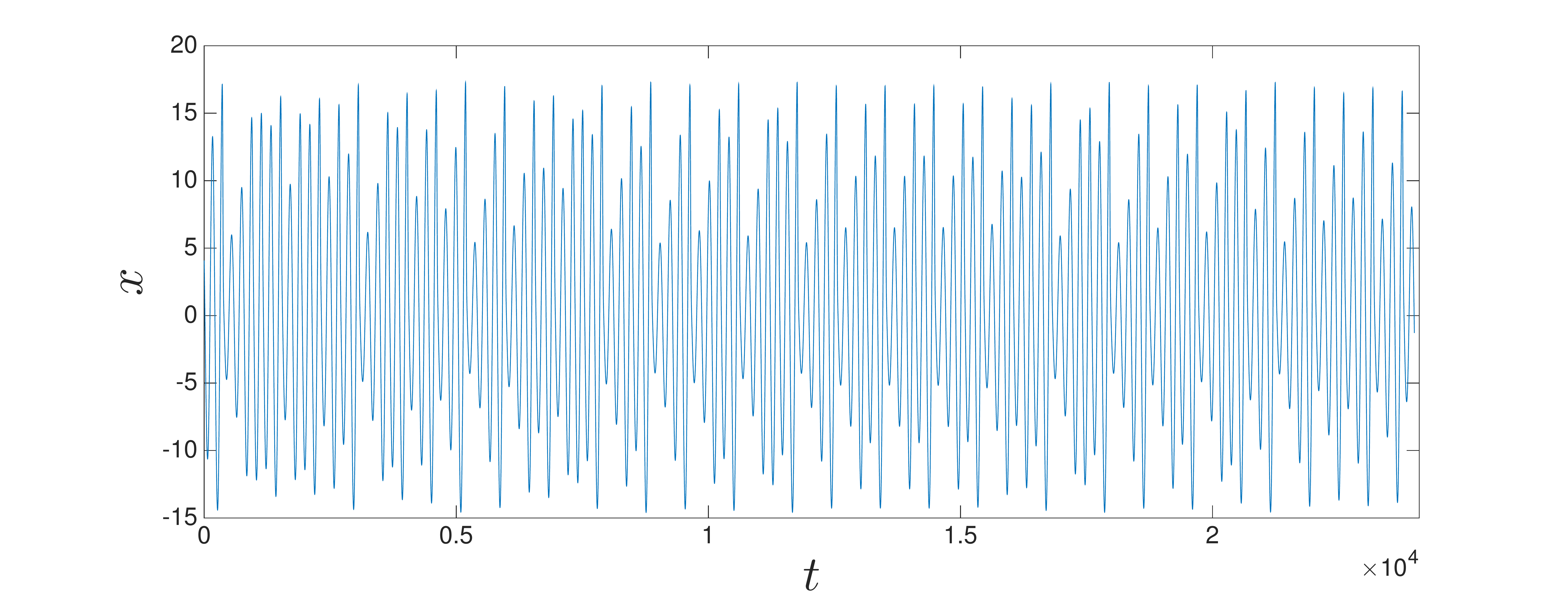}
    \includegraphics[width=1.8\columnwidth]{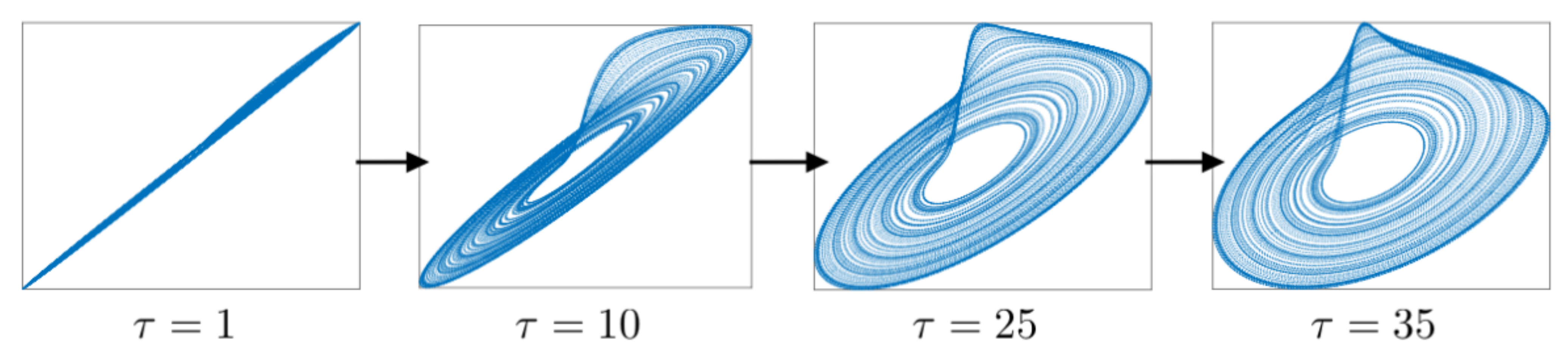}
    \caption{A time series from the Rossler system (top) and a number
of delay-coordinate embeddings of that time series with different
values of the delay parameter, $\tau$.}
   \label{fig:rossler}
  \end{figure*}

The standard strategy for state-space reconstruction is
delay-coordinate embedding, where a series of past values of a single
scalar measurement $y$ from a dynamical system are used to form a
vector that defines a point in a new space.  Specifically, one
constructs $m$-dimensional {\sl reconstruction-space} vectors
$\vec{R}(t)$ from $m$ time-delayed samples of the measurements
$y(t)$, such that \[\vec{R}(t) = [y(t), \; y(t -
\tau), \; y(t - 2 \tau), \; \dots \; , \; y(t - (m-1) \tau) ] \]
%
%
An example is shown in Figure~\ref{fig:rossler}. Mathematically, one
can equivalently take forward delays instead of backward ones, but for
practical purposes (e.g., predictions) it is better to obey causality
in one's notation.
If $\tau$ is very small, the $m$ coordinates in each of these vectors
are strongly correlated and so the embedded dynamics lie close to the
main diagonal of the reconstruction space; as $\tau$ is increased,
that reconstruction `unfolds' off that subspace.

The original embedding theorems only require that $\tau$ be nonzero
and not a multiple of any any orbit's period.  This is only valid,
however, when one is using real-valued arithmetic on an infinite
amount of noise-free data from perfect sensors.  In practice---when
noisy, finite-length time-series data and floating-point arithmetic
are involved---one needs a higher $\tau$ to properly unfold the
dynamics off the main diagonal.  The $\tau=1$ embedding in
Figure~\ref{fig:rossler}, for instance, will be indistinguishable from
a diagonal line if its thickness is smaller than the measurement noise level.
Since improperly unfolded reconstructions are not
topologically conjugate to the true dynamics, this is a real problem.
For this and other reasons, it can be a challenge to estimate good
values for the $\tau$ parameter, as described in more depth in
Section~\ref{sec:estimation}.

The original embedding theorems
also require $m>2d$ to assure topological conjugacy, where $d$ is the
true dimension of the underlying dynamics.  The trajectory crossings
in the two-dimensional projection of the embedded Rossler data in
Figure~\ref{fig:rossler}, for instance, do not exist in the real
attractor, and so the two structures do not have the same topology.
Sauer {\sl et al.} \cite{sauer91} loosened this requirement to $m>2
d_A$, where $d_A$ is the capacity dimension of the attractor.  In
practice, however, $d$ is rarely known and $d_A$ cannot be calculated
without first embedding the data.  A large number of heuristic methods
have been proposed to work around this quandary.  Many of these
methods are computationally expensive, most of them require
significant interpretation by---and input from---a human expert, and
all of them break down if one has a short or noisy time series.  These
methods, and their limitations, are also discussed in
Section~\ref{sec:estimation}.

There are two other requirements in the delay-coordinate embedding
theorems, one of which is implicit in the formula above: that one has
evenly spaced values of $y$.  Data-acquisition systems do not have
perfect time bases, so this can be a problem in practice.  An obvious
workaround here is interpolation, but then one is really studying a
mix of real and interpolated dynamics.  The final requirement is that
the measurement process that produces $y$ is a smooth, generic
function on the state space of the system.  This will not be the case,
for instance, if some event counter in the data-acquisition system
overflows.  It can be hard to know whether the measurement function satisfies
the theoretical requirement; strategies for doing so include changing the sampling
frequency or measuring a different quantity and then repeating the
analysis.  If the results do not change, one can be more confident
that they are correct.  Formal proofs of that correctness, of course,
are not possible because of the nature of real-world data and digital
computers.

Multivariate time-series data are useful for other reasons besides the
corroboration that is afforded via individual analyses of different
components.  It is also possible to perform {\sl multi-element}
reconstructions that combine the information in those components.  In
their 1980 paper\cite{packard80}, Packard {\sl et al.}  conjectured
that any $m$ quantities that ``...uniquely and smoothly label the
states of the attractor'' could serve as effective elements of a
reconstruction-space vector.  This powerful idea
%
%
is used surprisingly rarely, even though it is fully supported by all
routines of the {\sc tisean} software package\cite{tisean}.
%
%
With the improvement of sensor technology, the dynamical analysis of
multivariate data will likely become more important in the coming
years, as discussed in Section~\ref{sec:2039}.

The kind of `due diligence' exercise mentioned above is critical to
the success of any nonlinear time-series analysis task.  Data length,
noise, nonstationarity, algorithm parameters, and the like have strong
effects on the results, and the only way to know whether those effects
are at work in one's results is to repeat the analysis while
manipulating the data (downsampling, for instance, or analyzing the
first and last half of the data set separately) and the analysis
parameters---the $m$ and $\tau$ values, the algorithmic scale
parameters, etc.  If one can also manipulate the {\sl experimental}
parameters, repeated analyses can reveal whether the data are sampled
too coarsely in time to capture the details of the dynamics, or for
too short a period to sample its overall structure.

\subsection{Estimation of embedding parameters}
\label{sec:estimation}

The theoretical requirements on the embedding parameters---the delay
$\tau$ and the embedding dimension $m$---are, as mentioned in the
previous section, quite straightforward.  In practice, however, one
does not know the dimension of the system under study, nor does one
have perfect data or a computer that uses infinite-precision
arithmetic.  Estimating good value for $m$ and $\tau$ in the face of
these difficulties is one of the main challenges of delay-coordinate
embedding.  Dozens of methods for doing so have been developed in
the past few decades; we will only cover a few representative members
of this set.

In traditional practice, one chooses $\tau$ first, most often by
computing a statistic that measures the independence of
$\tau$-separated points in the time series.  The first zero of the
autocorrelation function of the time series, for instance, yields the
smallest $\tau$ that maximizes the linear independence of the
coordinates of the embedding vector; the first minima of the average
mutual information \cite{fraser-swinney} or the correlation
sum \cite{liebert-schuster89,grassberger-procaccia83} occur at $\tau$
values that maximize more-general forms of independence.  (One wants
the {\sl smallest} $\tau$ that is reasonable because the reconstructed
attractor can fold over on itself as $\tau$ grows, causing other
problems.)  There are also geometry-based strategies for estimating
$\tau$ by, for example, examining the continuity on the reconstructed
attractor or the amount of space that it fills.
While there has been some theoretical discussion\cite{casdagli91} of
what constitutes an optimal $\tau$, there are no universal strategies
for putting those ideas into practice---especially since the process
is system-dependent, and since a $\tau$ that works well for one
purpose (e.g., prediction) may not work well for another (e.g.,
computing dynamical invariants).

After choosing a value for $\tau$, the next step is to estimate the
embedding dimension $m$.  As in the case of $\tau$, bigger is not
necessarily better---since a single noisy point in the time series
will affect $m$ of the points in an $m$-embedding---so one wants the
smallest $m$ that affords a topologically correct result.  There are
two broad families of approaches to this, one based on the false near
neighbor (FNN) algorithm of Kennel {\sl et al.} \cite{KBA92} and
another that might be termed the ``asymptotic invariant'' approach.
In the latter, one embeds the data for a range of dimensions, computes
some dynamical invariant (e.g., those discussed in
Section~\ref{sec:invariants}), and selects the $m$ where the value of
that invariant settles down.  In an FNN-based method, one embeds the
data, computes each point's near neighbors, increases the embedding
dimension, and repeats the near-neighbor calculation.  If any of the
relationships change---i.e., some neighbor in $k$ dimensions is no
longer a neighbor in $k+1$ dimensions---that is taken as an indication
that the dynamics were not properly unfolded with $m=k$.  Noise also
disturbs neighbor relationships, though, and thus can affect the
operation of FNN-based algorithms.  No member of either family of
methods provides a guarantee, but both offer effective strategies for
estimating $m$.  Again, it can be very useful to employ several
different methods to corroborate one's results.

This two-step process is not the only approach.  It has been noted
that what really matters is the $m * \tau$ product---i.e., how much of
the data are spanned by the embedding vector---and thus that
estimating the two parameters at the same time, in combination, may be
more effective \cite{liebert-schuster91,pecora-attr-reconstr07}.  It
has also been suggested that one need not use the same $\tau$ across
all $m$ coordinates of the embedding vector---i.e., that a
systematically skewed embedding space may correspond better to the
true dynamics\cite{GrassbergerReview}.
%
%

\section{Mathematical beauty: 
Characterization of the invariant measure}\label{sec:invariants}


The invariant measure of a dynamical system can be characterized in a
number of different ways: the fractal dimension of the invariant set,
for instance, from the point of view of state-space geometry, or the
Kolmogorov-Sinai (K-S) entropy if one is interested in uncertainty
about the future of a chaotic trajectory.  The stability with respect
to infinitesimal perturbations can be quantified by the Lyapunov
exponents.  The topological equivalence guaranteed by the embedding
theorems allows all of these quantities---and many others not
mentioned here---to be determined from the time-series data.

\subsubsection{Dimension estimates}

There is a whole family of fractal dimensions $D_q$, usually called
the Renyi dimensions.  Their most intuitive definition is through a
partitioning of the state space: the number of boxes $N_\epsilon$ of
size $\epsilon$ needed to cover a fractal set with dimension $D_0$
scales with the box size $\epsilon$ as $\epsilon^{-D_0}$.  This is an
evident generalization of the integer dimensions, as one can easily
verify: a line segment, for instance, will yield $D_0=1$ via this
procedure, regardless of whether the surrounding space has two, three
or more dimensions.  $D_0$, often called the capacity dimension, is
closely related to the Hausdorff dimension\footnote{There is a
prominent exception to this statement: while the Hausdorff dimension
of the rational numbers is zero---as for any countable set of isolated
points---their capacity dimension is 1 because they are dense.}.  For
the generalized dimensions, one has to determine the measure on every
box from the partition and raise that measure to the power $q$, with
$\sum p_i^q \propto \epsilon^{(1-q)D_q}$ for $\epsilon\to 0$ and $p_i$
being the weight on the $i$th box.

Direct application of these box-counting methods to the points in the
reconstructed state space is possible, but involves significant memory
and processing demands and its results can be very sensitive to data
length.  A more-efficient, more-robust estimator of fractal dimensions
is the Grassberger-Procaccia correlation
sum\cite{GrassbergerProcaccia1981}.
%
%
We recall only the simplest version, which yields $D_2$.  Rather than
count boxes that are occupied by data points, one instead examines the
scaling of the {\sl correlation sum} as a function of $\epsilon$:
\begin{equation}
C_2(m,\epsilon) := \frac{1}{2N(N-T)}\sum_i\sum_{j<i-T} \Theta(\epsilon -
||\vec x_i-\vec x_j||)\; ,
\label{eq:corr-sum}
\end{equation}
where $\Theta$ is the Heaviside step function.  $C_2(m,\epsilon)$
represents the fraction of pairs of data points in the $m$-dimensional
embedding space whose spatial distance (measured by the Euclidean or
maximum norm) is smaller than the scale $\epsilon$.  This number
scales as $\epsilon^{D_2}$ if
$m>D_2$\cite{SauerYorke_refined_embedology}.  The parameter $T$, going
back to Theiler\cite{Theilerwindow}, ensures that the temporal spacing
between potential pairs of points is large enough to represent an
independently identically distributed sample\footnote{If $T$ is too
small, the estimate of $D_2$ might be biased towards too small
numbers, e.g., by intermittency effects.}.  

Formally, of course, the dimension of any finite point-set data should
be zero.  In the limit as $\epsilon \rightarrow 0$, methods that
simply count occupied boxes correctly reflect this fact.  In nonlinear
time-series analysis, however, we are interested in the dimension of
the set that is {\sl represented} by the point-set data.  The
correlation sum provides an unbiased estimator for that quantity, and
one that is accurate for small $\epsilon$---unlike the box method,
which is strongly biased towards small $D$ values in this
limit\cite{Grassberger_finite_sample}.

There is a conundrum involved in any estimation of the dimension of a
delay-coordinate embedding, which is sometimes known as the conflict
between redundancy and irrelevancy\cite{casdagli91}.  Specifically, in
order to assure that successive elements of a delay vector are
independent, the time lag $\tau$ should be sufficiently large.  This
can, however (as mentioned in the second paragraph of
Section~\ref{sec:estimation}) `overfold' the reconstructed
dynamics---especially if the embedding dimension is high.  In these
situations, it can require extremely well-sampled data in order to
correctly resolve the folds and voids in complicated chaotic attractors.
%
%
One can turn this reasoning around to estimate the number of data
points $N$ needed to estimate the dimension of a data set; a
pessimistic answer to this \cite{OlbrichKantz} is $N\approx
\sqrt{100^{D_2}e^{D_2h_2\tau}}$, where $h_{2}$ is the correlation
entropy of the dynamics, $\tau$ the time delay of the reconstruction,
and $e^{D_2h_2\tau}$ describes the effects of folding in the delay
embedding space due to the minimal embedding dimension $m>D_2$.  Among
other things, this means that the number of points needed to estimate
the dimension of chaotic dynamics reconstructed from a scalar time
series is much larger than in the original state space, where the
entropy factor can be ignored and $N>42^{D_2}$ has been
suggested\cite{Smith88}.

\subsubsection{Lyapunov exponents}

Dimension estimate have pitfalls and caveats, but they are quite
robust.  Estimates of Lyapunov exponents are unstable.  A number of
creative strategies have been developed for estimating the full set of
$m$ Lyapunov exponents $\lambda_k$ in the $m$-dimensional embedding
space (e.g.,~\cite{eckmann?}); there are also many algorithms for
estimating $\lambda_1$, the largest exponent, alone
(e.g.,~\cite{Wolf,SanoSawada}).  Every one of these algorithms
involves free parameters, however, and their results are often
extremely sensitive to the values of those parameters---as well as to
data length, noise, and the like.  When working with reconstructed
dynamics, one must also be aware of the issue of spurious exponents,
since the number of Lyapunov exponents is equal to the number of
dimensions in the ambient space.  Scalar time-series data sampled from
$D$-dimensional dynamical systems are typically embedded in $m$
dimensions with $m>D$, and those dynamics have $m$ Lyapunov exponents.
Ideally, one would like to find $D$ exponents that correspond to those
of the original dynamics---or at least to identify the $m-D$ extra
ones that are spurious.  There is a neat theory that predicts the
numerical values of these spurious exponents in lowest-order
approximation\cite{SauerYorke_spurious}, but this cannot usually be
reproduced in practice due to inaccessability of these
scales\cite{Hongliu+}.  

\subsubsection{The Kolmogorov-Sinai entropy}
\label{sec:ks}

Theoretically, the K-S entropy (rate) $h_{KS}$ can be determined via
Pesin's identity\cite{pesin77}, which states that it
is the sum of the positive Lyapunov exponents. Since spurious
exponents are hard to identify, though, and can even be positive, it
is difficult to put this into practice in the context of embedded data
(or to use the Kaplan-Yorke formula in order to determine the Lyapunov
dimension).  Rather, one typically estimates $h_{KS}$ through refined
partitions, closely following its definition (e.g.,~\cite{def_KS}).
The most straightforward implementation of this approach discretizes
the space of joint probabilities and searches for sequences of
successive delay vectors in specific sequences of boxes.  As in the
case of box-counting implementations of fractal dimension
calculations, this can lead to underestimation: a sequence that exists
in the underlying dynamics may not be `sampled' by a given set of
observations.  In the box-counting implementation, every sequence with
estimated probability 0 will systematically reduce the estimate of the
K-S entropy.  A way around this is to compute the correlation entropy
(rate) $h_2$, which can be estimated by the correlation sum.  To do
this, one calculates Eq.(\ref{eq:corr-sum}) for a range of dimensions
$m$ that are larger than the assumed minimum for an embedding,
obtaining $h_m(\epsilon)= \ln C(m,\epsilon)-\ln C(m+1,\epsilon)$.
Ideally, for some range of $\epsilon$, one should see a convergence of
$h_m(\epsilon)\to h_2$ for large $m$\footnote{$\epsilon$ values above
this range lead to underestimation; $\epsilon$ values below it lead to
large fluctuations.}.  For a consistency check, one can then go back
to Pesin's identity and compare the estimate of $h_2$ to the sum of
the positive Lyapunov exponents.  

\section{What practitioners need} \label{sec:practice}

A precise characterization of the invariant measure is not the goal of
most time-series analyses; moreover, few real-world data sets are
measured by perfect sensors operating on low-dimensional dynamics,
which means that a proper determination of, e.g., Lyapunov exponents,
is out of reach, anyway.  In practice, one typically wants to describe
a signal in some formalized manner, perhaps in order to discriminate
between it and some other signal.  Other important tasks include noise
reduction, detection of changes in dynamical properties within a given
signal, or prediction of its future values.  In all of these
situations, nonlinear time-series analysis has something to
contribute.

\subsection{Signal and system characterization}

A typical task is to characterize a single time series by a small set
of numbers---for the purposes of classification, for instance, or
comparison with other time series.  Examples include medical
diagnostics (is a patient healthy or sick?) or monitoring of machines
(is a lathe bearing wearing out?).  In these and many other important
applications, nonlinear time-series analysis offers a large zoo of
useful approaches, a few of which we describe below.

\subsubsection{Surrogate data}
\label{sec:surrogate}

In cases where strong evidence for some property is missing, one must
resort to statistical hypothesis testing.  With a finite data set, one
can never prove results about the underlying dynamics; one can only
calculate probabilities that a particular finding is unprobable using
a simple null hypothesis.  This approach can provide some evidence
that a more-complex (nonlinear, chaotic) dynamics is plausible, for
instance.

In nonlinear time-series analysis, the test statistics---Lyapunov
exponents, entropies, prediction errors, etc.---are complicated and
their probability distributions under simple null hypotheses are
typically unknown.  Furthermore, the ``simple'' null hypotheses are
typically not so simple.  In the face of these challenges, one can
proceed as follows.  First, one chooses a particular statistical
estimator (e.g., the violation of time inversion invariance\cite{SchreiberTI}, 
which is
a nonlinear property).  Second, one determines its value $v_d$ on the
target data set.  Third, one interprets that value by comparing it to
the distribution of values $v_s$ obtained from a large number of time
series that fulfill a certain null hypothesis (e.g., of AR processes).
Depending on where the computed value $v_d$ falls in this $v_s$
distribution, one can compute the probability of obtaining that value
``by chance.''
%
%
This provides a confidence level by which the null hypothesis can be
rejected.

How does one obtain the distribution of the test statistics under the
null hypothesis?  This is where {\sl surrogate data}\cite{surrogates1}
enter the game.
%
These are data that share certain properties of the time series under study and
also fulfill a certain null hypothesis.  The idea is that if one can
produce a number of such surrogate time series, one can numerically
compute the distribution of the test statistic on the null hypothesis.
The critical questions here are 
\begin{itemize}
\item which properties of the
original data should be shared by the surrogates?
\item what should the null hypothesis be?
\end{itemize}
Some of the answers are easy: since
insufficient time-series length poses severe problems, the individual
surrogate data sets should have the same length as the series under
study.  Others are not: ideally, for instance, each of these
sequences should represent the same marginal probability distribution
as the original data.  Since a rather powerful null model is the class
of ARMA models, it is reasonable to require the surrogate data to have
the same power spectrum (more precisely, the same periodogram) as the
original data---i.e., that temporal two-point correlations are
identical.  This is very useful when one wants to test for
nonlinearities, which express themselves in nontrivial temporal
$n$-point-correlations.


The technical way to create surrogate data with identical two-point
correlations and identical marginal distribution\cite{surrogates2}
is to Fourier transform one's original data, randomize the relative
Fourier phases, back transform (this creates close-to-Gaussian random
numbers with identical Fourier spectrum), and map the results onto the
original time-series values by rank ordering.  The third step restores
the original marginal distribution but partly destroys the
correlations,
%
so the power spectrum has to be re-adjusted by Wiener filtering.  Some
iteration of these steps is generally required until the features of
the surrogate data converge.  See \cite{schreiber_surrogates} for a
careful discussion of this family of methods.

While surrogate data tests are very useful---and very different than
the bootstrapping techniques used in other data-analysis
fields---there are a number of caveats of which one must be aware when
using them.  Prominent among these is the nonstationarity trap:
surrogates, by construction, are stationary, whereas the original data
may be nonstationary.  A difference in test statistics between
surrogates and original data, then, might have its origin in
nonstationarities rather than in nonlinearities.

\subsubsection{Permutation Entropy}
\label{sec:PE}

Since the 1950s, entropy has been a well-established measure of the
complexity and predictability of a time series\cite{Shannon1951}.
This is all very well in theory; in practice, however, estimating the
entropy of an arbitrary, real-valued time series is a significant
challenge.  The K-S entropy, for instance, is defined as the supremum
of the Shannon entropy rates of all partitions~\cite{petersen1989},
but not any partition will do for this computation.  There are
creative ways to work around this, as described in
Section~\ref{sec:ks}.  The main issue is discretization: these entropy
calculations require categorical data---symbols drawn from a finite
alphabet---but time-series data are usually real-valued and binning
real-valued data from a dynamical system with anything other than a
{\sl generating partition} can destroy the correspondence between the
true and symbolized dynamics\cite{lind95}.

{\sl Permutation entropy}\cite{bandt2002per} is an elegant way to work
around this problem.  Rather than computing the statistics of
sequences of categorical values, as in the calculation of K-S and
Shannon entropy, permutation entropy considers the statistics of
ordinal permutations of short subsequences of the time series.  If
$(x_1, x_2, x_3) = (9, 1, 7)$, for example, then its ordinal pattern,
$\phi(x_1, x_2, x_3)$, is $231$ since $x_2 \leq x_3 \leq x_1$.  The
ordinal pattern of the permutation $(x_1, x_2, x_3) = (9, 7, 1)$ is
$321$.  To compute the permutation entropy, one considers all the
permutations $\pi$ in the set $\mathcal{S}_n$ of all $n!$ permutations
of order $n$, determines the relative frequency with which they occur
in the time series, $\{x_t\}_{t = 1,\dots,T}$: $$ p(\pi)
= \frac{\left|\{t|t\le T-n,\phi(x_{t+1},\dots,x_{t+n})
= \pi\}\right|}{T-n+1}$$ where $|\cdot|$ is set cardinality, and
computes $$H_{PE}(n) = - \sum_{\pi \in \mathcal{S}_n}
p(\pi)\log_2p(\pi)$$ Like many algorithms in nonlinear time-series
analysis, this calculation has a free parameter: the length $n$ of the
subsequences used in the calculation.  The key consideration in
choosing it is that the value be large enough to expose forbidden
ordinal patterns but small enough that reasonable statistics over the
ordinals can be gathered from the given time series.  When this value
is chosen properly, permutation entropy can be a powerful tool; among
other things, it is robust to noise, requires no knowledge of the
underlying mechanisms, and is identical to the Shannon entropy for
many large classes of systems \cite{amigo2012permutation}.

\subsubsection{Recurrence plots}
\label{sec:RP}

\begin{figure}
    \centering
    \includegraphics[width=0.7\columnwidth]{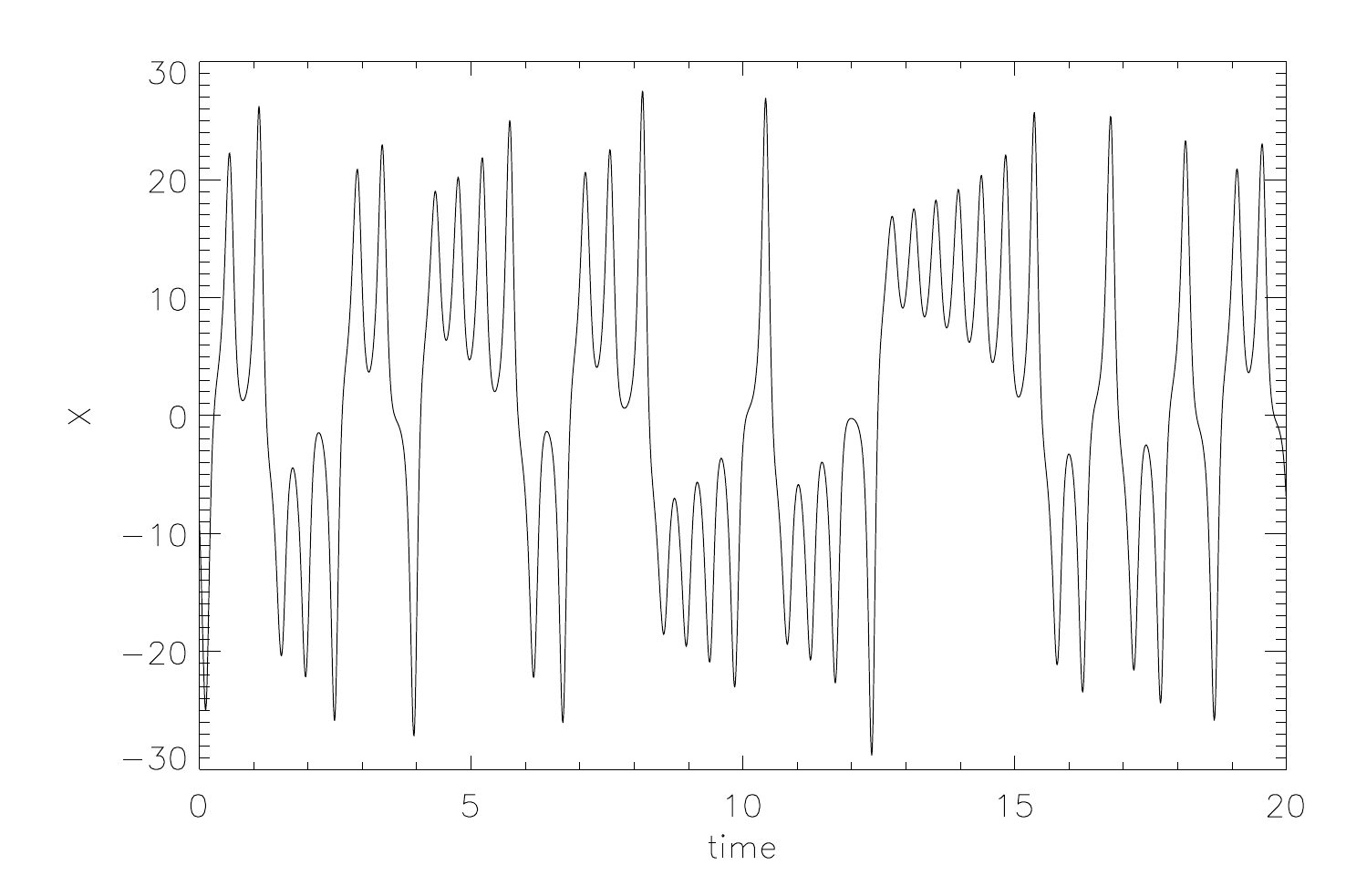}
    \includegraphics[width=0.7\columnwidth]{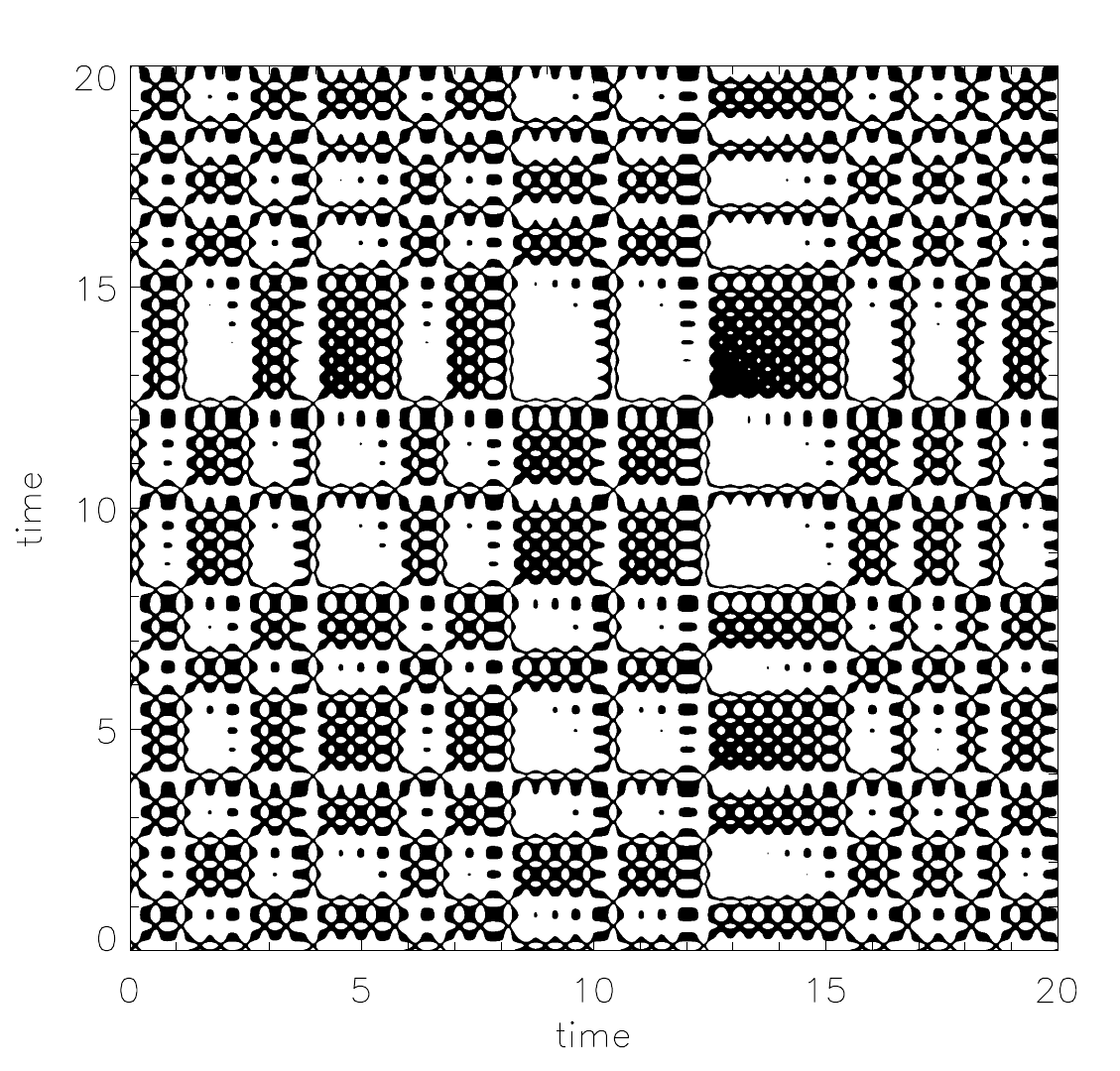}
\caption{\label{fig:rp} A signal and its recurrence plot.
Reproduced with permission from Chaos. 2:596 (2002). Copyright 2002 AIP Publishing.
%
%
}
\end{figure}

A recurrence plot\cite{eckmann87} is a two-dimensional visualization
of a sequential data set---essentially,
%
%
a graphical representation of the recurrence matrix of that sequence.
The pixels located at $(i,j)$ and $(j,i)$ on a recurrence plot (RP)
are black if the distance between the $i^{th}$ and $j^{th}$ points in
the time series falls within some {\sl threshold corridor}
$$\delta_l < || \vec{x}_i - \vec{x}_j || < \delta_h$$ for some
appropriate choice of norm, and white otherwise.  These plots can be
very beautiful, particularly in the case of chaotic signals; see
Figure~\ref{fig:rp} for an example.
(There are also ``unthresholded'' RPs, which use color-coding schemes
to represent a range of distances according to hue; these are even
more striking.)

RPs are useful in that they bring out correlations at all scales in a
manner that is obvious to the human eye, and they are one of the few
analysis techniques that work with nonstationary time-series data,
but their rich geometric structure---which, in the case of chaotic
signals, is related to the unstable periodic orbits in the
dynamics\cite{bradley-mantilla01}---can make them hard to interpret.
{\sl Recurrence quantification analysis} (RQA)\cite{zbilut92} defines
a number of quantitative metrics to describe this structure: the
percentage of black points on the plot, for example, or the percentage
of those black points that are contained in lines parallel to (but
excluding) the main diagonal.  RQA has been applied very successfully
to many different kinds of time-series data, notably from
physiological experiments (e.g., \cite{webber94}).  An extremely
useful review article is \cite{Marwan}.

\subsubsection{Network characteristics for time series}

Recently, recurrence plots have been interpreted in a very different
way, namely as the adjacency matrix of an undirected
network\cite{Donneretal}.  In this approach, an RP of an $N$-point
time series is converted into a network of $N$ nodes, pairs of which
are connected where the corresponding entries of the adjacency matrix
are non-zero.  One can then determine numerical values for different
network characteristics, such as centrality, shortest path length,
clustering coefficients, and many more.
%
%
There are some evident questions, the most relevant being about the
invariance of findings under variation of the threshold value
$\delta_l$,
%
since this value determines the link density of the network and all
network characteristics become trivial in the limit of full
connectivity.

\subsection{Prediction} \label{sec:prediction}

Prediction strategies that work with state-space models have a long
history---and a rich tradition---in nonlinear dynamics.  The
reconstruction machinery of Section~\ref{sec:basics} plays a critical
role in these strategies, as it allows them to be brought to bear on
the problem of predicting a scalar time series\cite{eckmann85}.  In
1969, for instance, Lorenz proposed his ``Method of Analogues,'' which
searches the known state-space trajectory for the nearest neighbor of
a given point and takes that neighbor's forward path as the
forecast\cite{lorenz-analogues}; not long after the original embedding
papers, Pikovsky showed that the Method of Analogues also works in
reconstructed state spaces\cite{pikovsky86-sov}.

Of course the canonical prediction example in deterministic nonlinear
dynamics is the roulette work of the Chaos Cabal at the University of
California at Santa Cruz,
a project that not only catalyzed a lot of nonlinear
science---including the original embedding paper\cite{packard80}---but
also a lot of interest in the field from both scientific and lay
communities\cite{bass} that continues to this day\cite{small12}.

In the decades since Lorenz's Method of Analogues and the roulette
project, a large number of creative strategies have been developed to
predict the future course of a nonlinear dynamical
system\cite{casdagli-eubank92}.  Most of these methods build some
flavor of local model in `patches' of a reconstructed state space and
then use that model to make the prediction of the next point.  Early
examples include \cite{farmer87,Smith88,casdagli89,sugihara90}.  This
remains an active area of research and has even spawned a time-series
prediction competition\cite{weigend-book}.

The Method of Analogues is not only applicable to deterministic
dynamics.  The short-term transition probability density of a Markov
process depends only on the current state, which can be approximated
by a delay vector.  The ``futures'' of delay vectors from a small
neighborhood can be viewed as a sample of the distribution, one time
step ahead.  This approach has been used for modeling\cite{Paparella}
and predicting\cite{Ragwitz} nonlinear stochastic processes.

Surprisingly, perfect embeddings are not required for successful
predictions.  In particular, reconstructions that do not satisfy the
theoretical requirements on the embedding dimension $m$ can give
prediction methods enough traction to match or even exceed the
accuracy of the same methods working in a full
embedding---particularly when the data are noisy \cite{josh-CHAOS15}.
One can then try to optimize, e.g., the embedding parameters.  Of
course, overfitting can be an issue in any prediction strategy; one
must be careful not to fool oneself by over-optimizing a predictor to
the given data.

\subsection{Noise and filtering} \label{sec:noise}

All real-world signals are contaminated by measurement noise.  Most
commonly, noise is treated as an additive random process on top of the
true signal.  Some forms of experimental apparatus contaminate the
signal in different ways, however: ``shot'' noise, for instance, which
appears only intermittently, or systematic bias in some measurement
device.  Regardless of its form, noise can interfere with nonlinear
time-series analysis if it is too large---where ``too large'' depends
greatly on the method that one wants to use.

Many studies in the literature are concerned with the fundamental
issue of distinguishing chaos and noise (see \cite{Chaosornoise} and
references therein).  This can be a real challenge.  Both types of
signals exhibit irregular temporal fluctuations, with a fast decay of
the auto-correlation function, and both are hard to forecast.  They
differ in the dynamical origin of these features: chaos is a
deterministic process, noise not.  In a deterministic system, the
short-term futures of two almost-identical states should be similar;
in a pure noise process that is improbable.  But, as mentioned above,
noise takes on many forms.  The simplest and most tractable is white
noise: sequences of independently identically distributed (iid) random
numbers.  Their statistical independence, as expressed by the
factorization of their joint probability distributions, can be easily
identified by statistical tests.

If the noise is not additive, the challenge mounts.  A noise-driven
chaotic system---e.g., a nonlinear stochastic differential
equation---produces something we might call noise.  Mathematically
speaking, such a system will, in any delay-coordinate embedding space,
generate an invariant measure whose support has the full state-space
dimension without fractal structure.  In such a system,
infinitesimally close trajectories will not diverge exponentially
fast, but rather separate {\sl diffusively}, at least on short time
scales.  Nonetheless, if such {\sl dynamical noise} or {\sl
  interactive noise} is sufficiently weak, one can still identify and
characterize the deterministic properties of the system.  However,
there is often a smooth transition between chaos and noise, leaving
the whole issue without a clear resolution.

It is, however, our impression that this issue is over-emphasized.  In
most time-series applications, it is not most critical to
``distinguish between chaos and noise,'' but rather to decide on the
complexity of the process: whether is it linear or nonlinear, where it
falls on the spectrum between redundancy and irrelevancy, etc.  And
then we are much better off, as there exist quite powerful tools for
answering these questions (see, e.g., Sections~\ref{sec:surrogate}
and~\ref{sec:PE}).

{\sl Removing} noise from a signal can also be a real challenge.
Traditional filtering strategies discriminate between signal and noise
using some sort of frequency threshold: e.g., removing all of the
high-frequency components of the signal.  In a chaotic signal, where
the frequency spectrum is broad band, such a scheme will filter signal
out along with the noise\cite{bleach}.  To be effective, filtering
strategies for nonlinear time-series data must be tailored to and
informed by the unique properties of nonlinear dynamics.  One can, for
instance, use the native geometry of the stable and unstable manifolds
in a chaotic attractor\cite{farmer88} or local models of the dynamics
on the attractor\cite{kostelich88,ghkss}, to reduce noise.  One can
also exploit the {\sl topology} of such attractors in nonlinear
filtering schemes\cite{robins-rooney-bradley02}.

\subsection{Issues and limitations} \label{sec:issues}

Nonlinear time-series analysis in the reconstructed state space is a
powerful and useful idea, but it does has some practical limitations.
These limitations are by no means fatal, but one has to be aware of
them in order to report correct results.

In theory, delay-coordinate embedding is only guaranteed to work for
an infinitely long noise-free observation of a {\sl single} dynamical
system.  This poses a number of problems in practice, beginning with
nonstationarity: embedding a time series gathered from a system that
is undergoing bifurcations, for instance, will produce a topological
stew of those different attractors.  ``Invariants'' computed from such
a structure, needless to say, will not accurately describe any of the
associated dynamical regimes.  One can use the tests described at the
end of Section~\ref{sec:dce} to determine whether these effects are at
work in one's results: e.g., repeating the analysis on different
subsequences of the data and seeing if the results change.  The
recurrence plots described in Section~\ref{sec:RP} can also be helpful
in these situations, allowing one to quickly see if different parts of
the signal have different dynamical signatures.

The analysis of different subsequences of a time series has many other
uses besides detecting nonstationarity, including determining whether
or not one has enough data to support one's analysis.  The original
embedding theorems require an infinite amount of data, but looser
bounds have since been established for different
problems\cite{Tsonis93,Smith88,eckmann92}.  It is important to know
and attend to these limits; a computation of a Lyapunov exponent of a
five-dimensional system from a data set that contains 100 points, for
instance, should probably not be trusted.  It is also important to
keep these effects in mind when repeating analyses on subsets of one's
data, since the changes in the results that one wants to use as a
diagnostic tool can simply be the result of short data lengths.

Dimension is a major practical issue for many reasons, not just
because it is not known {\sl a priori} and can be a challenge to
estimate.  Most of the results cited above regarding the data length
that is necessary for success in nonlinear time-series analysis scale
with the dimension of the dynamical system---often quite badly.  This
becomes even more of a challenge in spatially extended systems,
%
%
where the state space is high (or even infinite) dimensional and the
dynamics is spatio-temporal.  In cases like this, the full attractor
cannot be reconstructed by delay-coordinate embedding.  This can in
some cases be circumvented by exploiting homogeneity of the system,
however: if the dynamics is translationally invariant, local dynamics
can be reconstructed and used for predictions\cite{Baer,Parlitz}.

{\sl Noise effects} also scale with dimension, since any noisy
time-series point will affect $m$ of the points in an $m$-dimensional
embedding of those data.  The detection and filtering strategies
mentioned in Section~\ref{sec:noise} can help with noise problems, and
subsequence analysis can be used to explore whether the data are
adequate to support the analysis, but in the end there is simply no
way around not having enough data.

Delay-coordinate embedding, as formulated at the beginning of
Section~\ref{sec:basics}, requires data that are evenly sampled in
time.  If this is not true, constructing the delay vector $\vec{R}$ is
impossible without interpolation, which introduces spurious dynamics
into the results.  There is, however, an elegant way around this issue
if the data consist of discrete events, like the spikes generated by a
neuron: one simply embeds the inter-spike intervals\cite{sauer-isi95}.
The idea here is that if the spikes can be considered to be the result
of an integrate-and-fire process, then their spacing is an effective
proxy for the integral of the corresponding internal variable, and
that is a wholly justifiable quantity to embed.  Even without
integrate-and-fire dynamics, one can interpret interspike intervals
as a specific Poinar\'e map, which justifies their
embedding\cite{HeggerKantz97}.  This also applies to the time series
formed by all maxima (all minima) of the signal.

Even though for practical purposes it is quite handy, using the same
value of $\tau$ in between successive elements of a delay vector may
not be optimal.  Indeed, using delay vectors of the form
$y(t),y(t-\tau_1),y(t-\tau_1-\tau_2), ...., y(t-\tau_1-\tau_2-\ldots -
\tau_{m-1})$, with non-negative $\tau_i$, can introduce more time
scales into the reconstruction, which has been shown to be useful in
many situations\cite{pecora-attr-reconstr07,Holstein}.  Such
strategies might be also a way to tackle signals from multi-scale
dynamics: if there are different time and length scales involved, a
fixed $\tau$ may be too large to resolve the short ones and/or too small
to resolve the long ones.  This is particularly evident when embedding
a human ECG signal: using standard delay vectors, one can either
unfold the QRS complex or represent the t-wave as a loop, but not
both\footnote{Concerening spatial scales, it has been
shown\cite{Vulpiani_fsle} that spatial distances might play a
different role: so called finite-size Lyapunov exponents might detect
different strengths of instability of different spatial scales.}.

\section{Perspectives} \label{sec:2039}

When getting involved in time-series analysis some 25 years ago, we
could not have anticipated the wealth of data that would be available
in 2015, facilitated by cheap and powerful sensors for all sorts of
quantities, data-acquisition systems with sub-microsecond sampling
rates and terabytes of memory, widespread remote-sensing technology,
and incredible sense/compute power in small devices carried by the
majority of the population of Earth.  Commercial hardware and software
are available to monitor all kinds of things, from physiological
parameters obtained during daily activity by watch-sized objects to
real-time traffic flows gathered by cameras on highways.  These data
can be used to suggest life-changing health interventions, produce
routing suggestions to avoid traffic jams that have not yet formed,
and the like.

All this involves data analysis: often, time-series analysis.  The
bulk of the techniques used in the various academic and commercial
communities that are concerned with this problem---data mining,
machine learning, and the like---are linear and statistical.  Analysis
techniques that accommodate nonlinearity and determinism could be an
extremely important weapon in this arsenal, but nonlinear time-series
analysis is currently underused outside the field of nonlinear
science.  (Of course, much of this software is proprietary, so one
must be careful about such generalizations; nonlinear time-series
analysis may already be running on Google's computers and it would be
hard for those outside the company to know.)

There are some serious barriers for the movement of nonlinear
time-series analysis beyond the university desks of physicists and
into widespread professional practice, however.  Linear techniques
have a long history and are taught in most academic programs.  They
are comparatively easy to use and they almost always produce answers.
Whether or not those answers are correct, or meaningful, is a serious
issue: cf., the discussion in Section~\ref{sec:intro} of the mean of a
bimodal distribution.  But to a community that is familiar with these
linear techniques, the notion of learning a whole new
methodology---one that relies on more-complex mathematics and only
works if the data are good and the algorithm parameters are set
right---can be daunting.  One of us (EB) encountered significant
resistance when attempting to convince the computer systems community
to attend to nonlinearity and chaos in computer dynamics---an effect
that could significantly impact the designs of those systems.  Only
when those effects become apparent and meaningful to those communities
will nonlinear time-series analysis become more widespread.  Another
relevant issue here is whether low-dimensional deterministic dynamics
is a good data model for broader use.  So the only prediction that we
make here is that nonlinear time-series analysis is still far from its
culmination point, in terms of application.

What will be the relevant issues concerning the methodology itself?
Here we can only speculate.  It is evident that nonstationarity is
still a major problem and many of its facets are not fully explored.
Change-point detection is one of these.  Distilling causality
relationships from data is another critical open problem in nonlinear
time-series analysis (e.g., couplings in climate science).  Will this
ever be possible?  It is hard to say.  On the algorithmic end of
things, the various free parameters---and the sensitivity of the
results to their values---are important issues.  Will it be possible
to design algorithms whose free parameters can be chosen
systematically, via intuition, or perhaps even automatically?  Such
developments would streamline nonlinear time-series analysis, making
it an indispensible tool to make sense out of the real world.

\bibliography{references}

\end{document}